\documentclass[aps,eqsecnum,nofootinbib,showpacs]{revtex4}
\usepackage{amsmath,amsfonts,bm,amssymb}
\usepackage{color}

\newcommand{\be}{\begin{equation}}
\newcommand{\ee}{\end{equation}}
\newcommand{\bea}{\begin{eqnarray}}
\newcommand{\eea}{\end{eqnarray}}

\newcommand{\eq}[1]{Eq.~(\ref{eq:#1})}
\newcommand{\sect}[1]{Sec.~\ref{sec:#1}}
\newcommand{\appen}[1]{App.~\ref{sec:#1}}

\newcommand{\tabl}[1]{Table~\ref{table:#1}}

\newcommand{\nq}{\mathfrak{q}}
\newcommand{\nw}{\mathfrak{w}}
\newcommand{\del}{\partial}
\newcommand{\tpi}{\tau_{\pi}}
\newcommand{\tPi}{\tau_{\Pi}}
\newcommand{\tR}{\tau_J}
\newcommand{\fm}{{\rm fm}}

\bmdefine{\bmn}{ \bm{\rho} }
\bmdefine{\bmp}{ \bm{p} }
\bmdefine{\bms}{ \bm{s} }
\bmdefine{\bmP}{ \bm{k} }
\bmdefine{\bmT}{ \bm{T} }
\bmdefine{\bmU}{ \bm{U} }
\bmdefine{\bmeps}{ \bm{\epsilon} }
\bmdefine{\bmmu}{ \bm{\mu} }
\bmdefine{\bmnu}{ \bm{\nu} }
\newcommand{\hp}{ \Hat{p} }
\newcommand{\hs}{ \Hat{s} }
\newcommand{\hT}{ \Hat{T} }
\newcommand{\hmu}{ \Hat{\mu} }

\newcommand{\tilQ}{ \Tilde{Q} }

\newcommand{\vecx}{ \Vec{x} }
\newcommand{\invj}{J}
\newcommand{\invu}{U}

\bmdefine{\bmM}{ \bm{M} }

\begin{document}


%
\title{
Causal hydrodynamics of gauge theory plasmas from AdS/CFT duality}
\author{Makoto Natsuume}
\email{makoto.natsuume@kek.jp}
\affiliation{Theory Division, Institute of Particle and Nuclear Studies, \\
KEK, High Energy Accelerator Research Organization, Tsukuba, Ibaraki, 305-0801, Japan}
\author{Takashi Okamura}
\email{tokamura@kwansei.ac.jp}
\affiliation{Department of Physics, Kwansei Gakuin University,
Sanda, Hyogo, 669-1337, Japan}
\date{\today}
\begin{abstract}
We study causal hydrodynamics (Israel-Stewart theory) of gauge theory plasmas from the AdS/CFT duality. Causal hydrodynamics requires new transport coefficients (relaxation times) and we compute them for a number of supersymmetric gauge theories including the ${\cal N}=4$ SYM. However, the relaxation times obtained from the ``shear mode" do not agree with the ones from the ``sound mode," which implies that the Israel-Stewart theory is not a sufficient framework to describe the gauge theory plasmas.
\end{abstract}

\maketitle

\section{Introduction and summary}

The AdS/CFT duality is a powerful tool to study hydrodynamics of gauge theory plasmas, and it has interesting implications even to quark-gluon plasma (QGP). (See Refs.~\cite{Natsuume:2007qq,Son:2007vk,Mateos:2007ay} for reviews.) One robust prediction is  a universally small ratio of the shear viscosity $\eta$ to the entropy density $s$ at large 't~Hooft coupling:
\be
\frac{\eta}{s} =\frac{\hbar}{4\pi k_B}.
\label{eq:universality}
\ee
Similarly, one can compute the other transport coefficients such as bulk viscosity, speed of sound, and thermal conductivity.

However, standard hydrodynamics (first order formalism) has severe problems such as acausality. The first order formalism has the other problems: Equilibrium states are unstable under small perturbations \cite{hiscock_lindblom2} and the diffusion equation is inconsistent with sum rules  \cite{kadanoff_martin}. 

One can restore causality, but one is forced to introduce a new set of transport coefficients. Such a theory is known as ``causal hydrodynamics"  or  ``second order formalism." At present, there is no unique formalism for causal hydrodynamics. But probably the most used formalism is the ``Israel-Stewart theory" \cite{Israel:1976tn,Israel:1979wp}. (See Refs.~\cite{Muronga:2003ta,Maartens:1996vi} for reviews.) Another well-known candidate is the ``divergence type theories" \cite{LMR,Geroch:1990bw}, which has more attractive features mathematically. In this paper, we focus on the Israel-Stewart theory.

New coefficients which appear in the Israel-Stewart theory may become important in the early stage of QGP formation, and in fact it has been widely discussed in the context of heavy-ion collisions. For example, a number of groups recently reported the results of the $(2+1)$-dimensional numerical simulations of causal hydrodynamics \cite{Romatschke:2007mq,Chaudhuri:2007qp,Song:2007fn,Dusling:2007gi,Song:2007ux}. Unfortunately, little is known about these coefficients: They have been evaluated only for the Boltzmann gas (dilute gas approximation). 

The aim of this paper is to determine these coefficients from the AdS/CFT duality. The AdS/CFT duality cannot directly compute these coefficients for QCD, so we compute them for various supersymmetric gauge theories including the ${\cal N}=4$ SYM to see if there is any universality or generic features. We determine these coefficients by solving perturbation equations in the Schwarzschild-AdS black holes (SAdS) in various dimensions. 

Our results are summarized as follows:
\begin{enumerate}
%
\item One of the new transport coefficients is $\tpi$, which is the relaxation time for the shear viscous stress. This coefficient appears both in the ``shear mode" and in the ``sound mode," but their values do not coincide. This suggests that the Israel-Stewart theory is not sufficient to describe the gauge theory plasmas.
\item If one trusts the value of $\tpi$ obtained from the sound mode, %
\footnote{Our computations do not show which modes are reliable to obtain $\tpi$. However, just before we submitted the first version of this paper, Ref.~\cite{Baier:2007ix} appeared, which studies the similar problem as ours. They argue that the shear mode is unreliable to obtain $\tpi$, but the sound mode is reliable. Thus, we use our sound mode results for physical interpretations.}
$\tpi\sim 0.2$~fm (for $T^{-1} =1$~fm.) Using the AdS/CFT value of $\eta/s$, this value is not far from the Boltzmann gas estimation.%
\item The numerical values of $\tpi$ are similar among the theories we consider.
\end{enumerate}
Explicit results can be found in \sect{results}. In addition, we obtain the relaxation time $\tR$ for the charge diffusion in those theories.  
The coefficient $\tpi$ has been reported in Ref.~\cite{Heller:2007qt} for the ${\cal N}=4$ SYM using an expanding plasma. We compare our results and remark implications in an appropriate place (See also Ref.~\cite{Natsuume:2007tz}.)


In the next section, we illustrate the idea of causal hydrodynamics using a simple example. For the technical details used in this paper, see \appen{IS}. We set up perturbation equations in \sect{gravity} and present our results in \sect{results}.  

\section{Basic idea of causal hydrodynamics}


In this section, we review the idea of causal hydrodynamics using the charge diffusion example. Our discussion here is heuristic, but it serves a good starting point since the dispersion relation used in this paper in fact takes the same form as this simple example as shown in  \appen{dispersion}. Readers who are familiar with the idea of causal hydrodynamics may skip this section and may go to \appen{IS} directly for technical details.

The basic set of equations is the conservation law and the constitutive equation (Fick's law for the charge diffusion):
\be
\del_\mu J^\mu=0~, \qquad J_i= -D \del_i \rho~,
\ee
where $D$ is the diffusion constant. These two equations lead to the diffusion equation:
\be
\del_0 \rho - D \del_i^2 \rho = 0~.
\label{eq:diffusion}
\ee
The diffusion equation is parabolic, which does not satisfy causality. In fact, the propagator of \eq{diffusion} in $(1+1)$-dimensions is given by
\be
\rho \sim \frac{1}{\sqrt{4\pi Dt}}\exp\left(-\frac{x^2}{4Dt}\right)~,
\ee
which has a small but nonvanishing value even outside the lightcone $x>ct$.

In order to restore causality, one needs a hyperbolic equation such as the Klein-Gordon equation. The conservation equation must be true, so what is wrong is Fick's law. In fact, if $\del_i\rho=0$ for $t=0$, then Fick's law tells that the current vanishes immediately, {\it i.e.}, $J_i(t)=0$ for $t\geq0$. However, one expects that the current should die away in reality. In order to incorporate this effect, Fick's law may be modified as
\be
\tR \del_0 J_i + J_i= -D \del_i \rho~,
\label{eq:modified}
\ee 
where $\tR$ is a new transport coefficient. In this case, one obtains $J_i(t) = J_i(0) e^{-t/\tR}$. Thus, the parameter $\tR$ is the relaxation time for the charge current $J_i$.

The modified law with the conservation equation leads to the telegrapher's equation:
\be
\tR \del_0^2\rho + \del_0 \rho - D \del_i^2 \rho = 0~,
\label{eq:telegrapher}
\ee
which is a hyperbolic equation. The new term may become important at early time or for rapid evolution. Also, one can regard this as a higher order expansion of an effective theory. Hydrodynamics is just an effective theory with infinite number of parameters phenomenologically, so it is natural that new parameters arise.

A propagating solution $\rho \propto e^{-iwt+iqz}$ leads to the dispersion relation
\be
- \tR\, w^2 - i\,w + D\,q^2=0~.
\ee
The wave-front velocity can be estimated by taking the $q\rightarrow\infty$ limit: $v_{\rm \,front} = \sqrt{D/\tR}$.
Thus, the equation is consistent with causality if $v_{\rm \,front}<c$ (For large $w$, higher order terms may become important though). Let us also consider the opposite limit $q\rightarrow 0$. Then, the dispersion relation for the hydrodynamic pole (whose dispersion relation satisfies $w(q) \rightarrow 0$ as $q \rightarrow 0$) is written by
\be
w = - i \,D\,q^2 - i\,D^2\, \tR\, q^4 + O(q^6)~.
\label{eq:dispersion}
\ee
(Only one solution is compatible with the low-energy limit.)

Israel carried out a systematic analysis \cite{Israel:1976tn}, but the resulting constitutive equations are still complicated. We restrict the case of linear perturbations and decouple each modes. Hydrodynamic modes are decomposed as follows:
\bea
J_\mu &\rightarrow& \left\{   \begin{tabular}{l}
                                                    longitudinal mode (diffusive mode) \\
                                                    transverse mode
                                                  \end{tabular}
                                    \right. \nonumber \\
T_{\mu\nu} &\rightarrow& \left\{   \begin{tabular}{l}
                                                              longitudinal mode (sound mode) \\
                                                              transverse mode (shear mode) \\ 
                                                              transverse traceless mode 
                                                            \end{tabular}
                                    \right. \nonumber
\eea
Not all modes have a hydrodynamic pole since such a pole arises due to a conservation law. The standard transport coefficients appear in the following modes: the charge diffusion constant $D$ in the diffusive mode (as is clear from the above example), the shear viscosity $\eta$ in the shear mode, the bulk viscosity $\zeta$ and the speed of sound $v_s$ in the sound mode. 

In addition, Israel introduced 5 new transport coefficients: Three are relaxation times for the diffusive, shear, and sound mode, respectively ($\tR, \tpi, \tau_\Pi$, respectively). The other two are the couplings among different modes (a coupling between the diffusive and the sound mode $\alpha_0$, and a coupling between the diffusive and the shear mode $\alpha_1$).  At the end of the day, {\it the dispersion relations for the diffusive and the shear mode just take the form (\ref{eq:dispersion}) for the telegrapher's equation.} [See Eqs.~(\ref{eq:hydro-decouple-dipersion_rel-vector}) and (\ref{eq:hydro-decouple-dipersion_rel-charge}) in \appen{dispersion}.] We determine these coefficients from the gravity computation below.

\section{Gravity computations}\label{sec:gravity}

According to the standard AdS/CFT dictionary, the bulk gauge field $A_\mu$ acts as the source for the global R-charge current on the dual field theory. (The R-charge is a global charge which presents in SYM; In this sense, it is an analog of the baryon number in QCD.) Similarly, the bulk gravitational perturbations act as the source for the stress-energy tensor on the dual theory. Thus, our aim is to solve these bulk field equations. 

Let us consider the bulk perturbations of a $p$-brane which take the form 
\be
A_{\mu}(r) \, e^{-iwt + iqz}~, \qquad
h_{\mu\nu}(r) \, e^{-iwt + iqz}~,
\ee
where $z:=x^p$. The perturbations can be decomposed by the little group $SO(p-1)$ acting on $x^i (i = 1, \cdots, p-1)$. The gauge field perturbations are decomposed as 
\bea
&& \mbox{scalar mode (diffusive mode): } A_0,\, A_z,\, A_r,  \nonumber \\
&& \mbox{vector mode: } A_i. \nonumber 
\eea
Similarly, the gravitational perturbations are decomposed as 
\bea
&& \mbox{scalar mode (sound mode): } h_{tt},\, h_{tz},\, h_{zz},\, h^k_{~k},\, h_{rr},\, h_{tr},\, h_{zr},  \nonumber \\
&& \mbox{vector mode (shear mode): } h_{ti},\, h_{zi},\, h_{ri}, \nonumber \\ 
&& \mbox{tensor mode: } h_{ij}-\delta_{ij} h^k_{~k}/(p-1).\nonumber 
\eea
Such a decomposition is essentially the same as hydrodynamics above. 
Each hydrodynamic mode couples to the corresponding bulk perturbation.

Many authors solve such perturbation equations in various backgrounds.%
\footnote{
For example, at the lowest order in $w$ and $q$, the perturbation equations have been first solved in Refs.~\cite{Policastro:2002se,Policastro:2002tn} for SAdS$_5$ and in Refs.~\cite{Herzog:2002fn,Herzog:2003ke} for SAdS$_{4,7}$. See also Ref.~\cite{Hartnoll:2007ip} for a recent application of SAdS$_4$.}
Our aim is to get the subleading corrections by regarding causal hydrodynamics as an effective theory expansion in higher orders. In this paper, we consider the diffusive mode, the shear mode, and the sound mode. 

\subsection{Backgrounds}\label{sec:backgrounds}

In this paper, we compute transport coefficients for the SAdS$_{p+2}$ backgrounds. These backgrounds appear as the ``near-horizon" limit of various branes. The $p=3$ case corresponds to the D3-brane which is the ${\cal N}=4$ SYM. The $p=2$ case corresponds to the D1-brane or M2-brane in 11-dimensional supergravity. The $p=5$ case corresponds to the D4-brane or M5-brane. The D1 and D4-branes are dual to the 2 and 5-dimensional SYM with 16 supercharges.%
\footnote{For a recent discussion of this duality, see, {\it e.g.},
Ref.~\cite{Maeda:2005cr} and references therein.}

The ${\rm SAdS}_{p+2}$ metric is given by
\be
ds_{p+2}^2 = 
f (-hdt^2+d\vec{x}_p^2) +  \frac{dr^2}{fh}~,
\label{eq:sads_metric}
\ee
where 
\bea
f &=& \left( \frac{r}{R} \right)^2~, \\
h &=& 1-\left( \frac{r_0}{r} \right)^{p+1}~.
\eea 
The surface gravity is given by
\be
\kappa=\frac{p+1}{2}\frac{r_0}{R^2}~.
\ee

Since some of the backgrounds can be interpreted as the D$p$-brane, let us directly consider the D$p$-brane for completeness.
The D$p$-metric consists of a $(p+2)$-dimensional metric and warped $S^{8-p}$. According to Ref.~\cite{Kovtun:2003wp}, the dimensional reduction of the metric into the $(p+2)$-dimension gives
\bea
ds_{p+2}^2 &=& 
r^{\frac{2(8-p)}{p}} 
\left\{ Z^{\frac{1}{p}} (-hdt^2+d\vec{x}_p^2) + Z^{1+\frac{1}{p}}  \frac{dr^2}{h} \right\}~,
\label{eq:dp_metric} \\
g_{\rm eff}^2 &=& r^{-\frac{16}{p}}Z^{-1-\frac{1}{p}}~,
\eea
where 
\bea
Z &=& \left( \frac{r}{R} \right)^{-(7-p)}~, \\
h &=& 1-\left( \frac{r_0}{r} \right)^{7-p}~.
\eea 
The surface gravity is given by
\be
\kappa = \frac{7-p}{2}\frac{r_0^{\frac{5-p}{2}}}{R^{\frac{7-p}{2}}}~.
\ee

\subsection{Field equations (diffusive and shear mode)}

Our computation closely follow Ref.~\cite{Kovtun:2003wp}. Let us start from the diffusive mode. This amounts to solve the Maxwell equation:
\be
\nabla_\mu(\sqrt{-g} F^{\mu\nu})=0
\label{eq:maxwell}
\ee
where $\sqrt{-g}:=\sqrt{-g_{p+2}}/g_{\rm eff}^2$. The effective coupling $g_{\rm eff}$ may be position-dependent.

It is convenient to introduce a new radial coordinate $u$ ($u:=r_0/r$ for even $p$ and $u:=r_0^2/r^2$ for odd $p$). We choose the gauge $A_u=0$ and use a Fourier decomposition:
\be
   A_{\mu}(u, t, z) = \int\! \frac{dw\, dq}{(2\pi )^2}\, 
   e^{-iw t + i q z}
   A_{\mu} (u, w, q)\,.
\label{eq:fourier}
\ee
Then, the Maxwell equation becomes
\bea
& & g^{00} w A_0' - q g^{zz} A_z' =0~,
\label{eq:maxwell1} \\
& & 
\partial_u \left( \sqrt{-g} g^{00} g^{uu} A_0'\right)
 -  \sqrt{-g} g^{00} g^{zz} \left( w q A_z + q^2 A_0 \right) =0~,
\label{eq:maxwell2} \\
& &
\partial_u \left( \sqrt{-g} g^{zz} g^{uu} A_z'\right) 
 -  \sqrt{-g} g^{00} g^{zz} \left( w q A_0 + w^2 A_z \right) =0~,
 \label{eq:maxwell3}
\eea
where  $~{}' = \partial_u$.
From Eqs.~(\ref{eq:maxwell1}) and (\ref{eq:maxwell2}), one gets a decoupled equation for $A_0'$:
\be
\frac{d}{du} \left[ 
\frac{\partial_u ( \sqrt{-g} g^{00} g^{uu} A_0')} {\sqrt{-g} g^{00} g^{zz}}
\right] 
+ \left(-\frac{g^{00}}{g^{zz}} w^2 - q^2\right) A_0' =0\,.
\label{eq:master1}
\ee
In the backgrounds we use, the equations of motion reduce to the following form: 
\be
\frac{d}{du}\left[ u^\alpha h \frac{d}{du}(u^\beta A_0') \right]
+\frac{\nu^2}{4}\left(\frac{\nw^2}{h}-\nq^2\right)A_0'=0~,
\label{eq:master2}
\ee
where $h=1-u^\nu$; $\alpha$, $\beta$, and $\nu$, are the constants which depend on the backgrounds; $\nw$ and $\nq$ are $w$ and $q$ normalized by surface gravity $\kappa$ [or temperature $T:=\kappa/(2\pi)$]:
\be
\nw=\frac{w}{2\pi T}, \qquad \nq=\frac{q}{2\pi T}~.
\ee

Incorporating the ``incoming wave" boundary condition at the horizon $u=1$ and asymptotic form 
$u=0$, 
\be
A_0'=C(1-u^\nu)^{-i\nw/2} u^{-\beta} F(u)~.
\ee
The function $F(u)$ is a regular function whose form can be obtained perturbatively as a double series in $\nw$ and $\nq^2$:
\be
F(u) = F_0 + \nw F_1 + \nq^2 G_1 + \nw^2 F_2 + \nw\nq^2H_{11} + \nq^4 G_2 +\cdots~.
\ee
The dispersion relation is obtained by imposing Dirichlet boundary condition at $u=0$. Such a dispersion relation has been obtained for various theories at $O(\nw, \nq^2)$. Our task is to compute corrections at $O(\nw^2, \nw\nq^2, \nq^4)$ to get the coefficients of causal hydrodynamics.

For the shear mode, denote one of $x^i$ coordinates as $x$. We consider a metric perturbation of the form $h_{tx}\neq0, h_{zx}\neq0$ with the other $h_{\mu\nu}=0$. As explained in Ref.~\cite{Kovtun:2003wp}, the equation for the shear mode reduces to the Maxwell equation. 
First, consider a fictitious Kaluza-Klein compactification along the $x$-direction. Following the standard procedure of the Kaluza-Klein reduction, set $A_0=(g_{zz})^{-1} h_{tx}$ and $A_z=(g_{zz})^{-1} h_{zx}$. Then, write the resulting action in terms of the Einstein metric.  In the end, the perturbation equation becomes \eq{master1} with the replacement
\be
\sqrt{-g} \rightarrow g_{zz} \sqrt{-g}~,
\label{eq:gravityASmaxwell}
\ee
and the only differences are the parameters $\alpha$ and $\beta$ in \eq{master2}.

Using the backgrounds in \sect{backgrounds}, one obtains $\alpha$, $\beta$, and $\gamma$ as shown in \tabl{parameters}. It is easy to see that the perturbation equations for the D1 and the D4-brane are identical to those for  SAdS$_4$ and SAdS$_7$, respectively. 

\begin{table}
\begin{center}
\begin{tabular}{|c||c|c|c||c|c|c|}
\hline
Geometry
& \multicolumn{3}{c||}{shear mode}
& \multicolumn{3}{c|}{diffusive mode} \\
\cline{2-7} 
& $\alpha$ & $\beta$ & $\nu$ & $\alpha$ & $\beta$ & $\nu$ \\
\hline
&&&&&& \\
SAdS$_{p+2}$ ($p$: even)
     & $ p $
     & $ -p $
     & $ p+1 $ 
     & $ p-2 $
     & $ 2-p $
     & $ p+1 $ \\
&&&&&& \\
SAdS$_{p+2}$ ($p$: odd)
     & $ \displaystyle{\frac{p+1}{2}} $
     & $ \displaystyle{-\frac{p-1}{2}} $
     & $ \displaystyle{\frac{p+1}{2}} $
     & $ \displaystyle{\frac{p-1}{2}} $
     & $ \displaystyle{\frac{3-p}{2}} $
     & $ \displaystyle{\frac{p+1}{2}} $ \\
&&&&&& \\
D$p$ ($p$: even)
     & $ 3 $
     & $ p-6 $
     & $ 7-p $ 
     & $ p-2 $
     & $ -1 $
     & $ 7-p $ \\
&&&&&& \\
D$p$ ($p$: odd)
     & $ 2 $
     & $ \displaystyle{\frac{p-5}{2}} $
     & $ \displaystyle{\frac{7-p}{2}} $
     & $ \displaystyle{\frac{p-1}{2}} $
     & $ 0 $
     & $ \displaystyle{\frac{7-p}{2}} $ \\
&&&&&& \\
\hline
\end{tabular}
\caption{The parameters $\alpha, \beta$, and $\nu$ appeared in \eq{master2}.}
\label{table:parameters}
\end{center}
\end{table}

\subsection{Field equations (sound mode)}

We closely follow Ref.~\cite{Kodama:2003jz}. 
The sound mode has 7 degrees of freedom. Out of these, there are 3 gauge freedoms and 3 constraints, which leaves us a single degree of freedom. Our task is to constitute a master field which represents this degree of freedom and to obtain the master equation.

First, let us compare our notations with those of Kodama and Ishibashi~\cite{Kodama:2003jz}. They take the $\text{SAdS}_{p+2}$ metric as
\begin{align}
  & ds^2
  = - f(r)\, dt^2 + \frac{dr^2}{f(r)} + r^2\, d\sigma_p^2
  = g_{ab}(y)\, dy^a dy^b + r^2(y)\, d\sigma_p^2~,
& & f(r) = - \lambda\, r^2 - \frac{2 M}{r^{p-1}}~,
\end{align}
where $(a, b) = (t, r)$. As a result, our notations and theirs are related as follows:
\begin{center}
\begin{tabular}{ccc} \hline
  \text{Kodama and Ishibashi} & \hspace{2.0cm} & \text{Ours}
\\ \hline
  $\lambda$ & & $- 1/R^2$
\\
  $M$ & & $r_0^{p+1}/2 R^2$
\\
  $f(r)$ & & $f(r)\, h(r)$
\\
  $d\sigma_p^2$ & & $d\vecx_p\!{}^2/R^2$
\\
  $k$ & & $R\, q$
\\ \hline
\end{tabular}
\end{center}
where $k$ and $q$ represent wave numbers in their and our notations, respectively.

They have written down the equations for 3 gauge-invariant variables $X(r)$, $Y(r)$, and $Z(r)$ [Eqs.~(2.24a)-(2.24d) in Ref.~\cite{Kodama:2003jz}]:
\begin{align}
   \frac{dX}{dr}
  &= \frac{p-2}{r}\, X + \frac{1}{h}\, \frac{dh}{dr}\, Y
  + \left( \frac{R^2}{r^2}\, \frac{q^2}{f\, h}
     - \frac{w^2}{(f\, h)^2} \right) Z~,
 \label{eq:2_24a} \\
   \frac{dY}{dr}
  &= \frac{1}{2 (f\, h)}\, \frac{d(f\, h)}{dr}\, ( X - Y )
  + \frac{w^2}{(f\, h)^2}\, Z~,
 \label{eq:2_24b} \\
   \frac{d Z}{dr}
  &= X~,
 \label{eq:2_24c} \\
   0
  &= \alpha(r)\, X + \beta(r)\, Y + \gamma(r)\, Z~,
 \label{eq:2_24d} 
\end{align}
where we define $\alpha$, $\beta$, and $\gamma$  by
\begin{align}
   \alpha(r)
  &:= w^2\, r^2 + \frac{p+1}{2}\, \left( \frac{r_0}{R} \right)^4
  \left( \frac{r_0}{r} \right)^{p-3}
    \left\{ p - \frac{p-1}{2}\, \left( \frac{r_0}{r} \right)^{p+1}
    \right\}~,
\label{eq:def-alpha} \\
   \beta(r)
  &:= w^2\, r^2 - q^2\, r^2\, h
  + \left( \frac{p+1}{2} \right)^2\, \left( \frac{r_0}{R} \right)^4\,
      \left( \frac{r_0}{r} \right)^{2(p-1)}~,
\label{eq:def-beta} \\
   \gamma(r)
  &:= - r\, \left[~p\, w^2 - q^2\,
     \left\{ 1 + \frac{p-1}{2}\, \left( \frac{r_0}{r} \right)^{p-3}
     \right\}~\right]~.
\label{eq:def-gamma}
\end{align}
Here, $X$, $Y$, and $Z$ are Fourier-transformed as in \eq{fourier}.
Equations~(\ref{eq:2_24a})-(\ref{eq:2_24c}) are coupled first-order differential equations for 3 variables, but they reduce to a second-order differential equation for a single variable by a constraint (\ref{eq:2_24d}), which is the master equation. 


However, the master equation derived by Ref.~\cite{Kodama:2003jz} [Eq.~(3.5) in their paper] is not particularly useful for our purpose. In order to solve the eigenvalue problem as a series in $\nw$ and $\nq$, it is necessary that one can take the limit $\nw,\, \nq \rightarrow 0$ not only for the perturbation equation but also for the boundary condition. Namely, suppose that the lowest-order solution as a series in $\nw$ and $\nq$ has some fall-off behavior as  $r \rightarrow \infty$. This fall-off behavior must coincide with the one for the full-order solution. 
Unfortunately, the master equation derived in Ref.~\cite{Kodama:2003jz} does not satisfy this criterion, so one must use a new master field and obtain the master equation for such a field for which one can take the limit $\nw,\, \nq \rightarrow 0$.

After some trial and error, we found that the following form of the master field $\Phi$ is useful:
\begin{align}
  & \Phi(s)
  := \frac{s^{p-2}}{h(s)}\, \left[~
     \left( 1 - \frac{p-1}{2\, p}\, s^{p+1} \right)\, X(s)
  + \frac{p+1}{2\, p}\, s^{p+1}\, Y(s)~\right]~,
\label{eq:def-Phi}
\end{align}
where 
\begin{align}
  & s := r_0/r~.
\end{align}
From Eqs.~(\ref{eq:2_24a})-(\ref{eq:2_24d}), the equation for $\Phi$ is schematically written as
%
%
%
\begin{align}
   0 = \frac{d^2 \Phi}{du^2}
 + B_1(u) \frac{d \Phi}{du}
 + B_0(u) \Phi~.
\label{eq:EOM-Phi}
\end{align}
For even $p$, 
\bea
B_1(u) &:=& 
  - \frac{2 p \big\{ p + (2p+3) u^{p+1} \big\}\, \nw^2
        + \nq^2 \big\{ (p-1) u^{2(p+1)}
                - (p^2+7p+2) u^{p+1} - 2p^2 \big\} }
         {u\, h\, \big[~2p\, \nw^2 + \nq^2
            \{ (p-1) u^{p+1} - 2 p \}~\big] }~,
\\
B_0(u) &:=& 
  \frac{(p+1)^2}{4}\, 
  \frac{1}{h^2\, \big[~2p\, \nw^2 + \nq^2 \{ (p-1) u^{p+1} - 2 p \}~\big] }
\nonumber \\
&& \times \big[ 8 p\, \nw^2 u^{2p} - 4(p+1) \nq^2 u^{2p}
       + 2 p\, \nw^4 + \nq^4 \big\{ (p-1) u^{2(p+1)}
                 - (3p-1) u^{p+1} + 2p \big\} 
\nonumber \\
&&      - \nw^2 \nq^2 \big\{ 4p - (3p-1) u^{p+1} \big\} \big]~,
\eea
where $ u:=s $.
For odd  $p~(= 2\, p'+1)$, 
%
%
\bea
B_1(u) &:=& 
  - \frac{ p\, \big\{ (p+2) u^{p'+1} + p' \big\}\, \nw^2
        + \nq^2 \big\{ p'\, u^{2(p'+1)}
                - (p'^2+6p'+3) u^{p'+1} - p\, p' \big\} }
         {u\, h\, \big\{ p\, \nw^2 + \nq^2
            ( p'\, u^{p'+1} - p ) \big\} }~,
\\
B_0(u) &:=& 
  \frac{(p'+1)^2}{4}\,
  \frac{1}{u\, h^2\, \big\{ p\, \nw^2 + \nq^2 ( p'\, u^{p'+1} - p ) \big\} }
\nonumber \\
&&   \times \big[ 4 p\, \nw^2 u^{p} - 4(p'+1) \nq^2 u^{p}
       + p\, \nw^4 + \nq^4\, h\, ( p - p'\, u^{p'+1} )
\nonumber \\
&&       - \nw^2 \nq^2 \big\{ 2 p - (3p'+1) u^{p'+1} \big\} \big]~,
\eea
where $u:=s^2$.
Finally, incorporating the boundary condition at the horizon $u=1$ and asymptotic form $u=0$, set 
\be
\Phi = h^{-1-i\nw/2} s^{p+1} F(u)~.
\label{eq:bc_sound}
\ee

For SAdS$_5$, the master equation for the sound mode has been obtained by Kovtun and Starinets \cite{Kovtun:2005ev} in the way consistent with the hydrodynamic limit [Eq.~(4.35) in the paper]. The above master equation is identical to theirs. Also, for the D$p$-brane, the master equation has been obtained by Mas and Tarrio \cite{Mas:2007ng} [Eq.~(3.17) in the paper]. The above master equation for SAdS$_4$ and SAdS$_7$ is identical to those for the D1 and the D4-brane, respectively. Our variable $F$ is related to the variables by the others  as follows:
\bea
\mbox{Kovtun and Starinets: \quad} 
Z_2(u) &=& h^{-i\nw/2} u^2 F(u)~, \\
\mbox{Mas and Tarrio: \quad} 
Z_0(u) &=& h^{-i\nw/2} F(u)~,
\eea
where $Z_2$ and $Z_0$ are the variables used in the papers above.

In general, the perturbation equations in the sound mode are harder to solve at the second order than the ones in the shear mode. Moreover, the generic SAdS$_{p+2}$ case is harder than the SAdS$_5$ case. In order to simplify our analysis, we employ the following method. First, anticipating the hydrodynamic dispersion relation, set
\be
\nw = d_0 \nq + d_1 \nq^2 + d_2 \nq^3 +\cdots~,
\ee
and obtain the solution as a series in $\nq$:
\be
F(u) = F_0 + \nq F_1 + \nq^2 F_2 + \cdots~.
\ee
The constant $d_i$ is obtained at each order by imposing Dirichlet boundary condition on the solution $F_i$. Then, we solve the equation for $F_{i+1}$ using $F_i$ (with the determined constant $d_i$).  

\section{Results}\label{sec:results}

\subsection{The shear mode and the diffusive mode}

The solutions $F(u)$ are rather cumbersome expressions, so we do not write them explicitly. (The solution for the shear mode and the sound mode of the ${\cal N}=4$ SYM are written in \appen{solutions}.) 
The Israel-Stewart theory has 5 new constants $(\tR, \tpi, \tPi, \alpha_0, \alpha_2).$ For the backgounds with no R-charge, the gauge field and the metric perturbations decouple: This implies that $\alpha_0=\alpha_1=0$. In addition, for conformal theories $\tPi=0$ due to the vanishing $\zeta$. [See \eq{def-tau_Pi}.] Thus, the main interests are $\tR$ and $\tpi$.

From the shear mode of the ${\cal N}=4$ SYM, we get
\be
\nw = -i\,\frac{\nq^2}{2} - i\,\frac{1-\ln2}{4} \nq^4 + O(\nq^6)~.
\label{eq:sads5_vector}
\ee
Comparing \eq{sads5_vector} with the dispersion relation (\ref{eq:hydro-decouple-dipersion_rel-vector}), we obtain the familiar result $\eta/s=1/(4\pi)$ and
\be
\tpi = \frac{1-\ln2}{2\pi T}~.
\label{eq:tau_pi_vector}
\ee
The other results are summarized in \tabl{resultsI} and \tabl{resultsII}. For the diffusive mode, one gets the diffusion constant $D$
\be
D = \frac{p+1}{4(p-1)\pi T}
\ee
as well as $\tR$.

For the ${\cal N}=4$ SYM, $\tR$ has never been obtained, but the result is obvious {\it a priori} from a result of Ref.~\cite{Policastro:2002se} and our dispersion relation for the diffusive mode (\ref{eq:decouple-dipersion_rel-charge}). Also, Ref.~\cite{Heller:2007qt} has computed $\tpi$ for the ${\cal N}=4$ SYM from a somewhat different setting. They consider an expanding plasma and obtained $\tpi$ which is 3 times smaller than our result. 
We believe that this is due to a missing term in their constitutive equation.  The term is negligible for a plasma near equilibrium, but it is not negligible for the expanding plasma,  In fact, the discrepancy is gone once one adds the extra term in the constitutive equation (See Ref.~\cite{Natsuume:2007tz} for details.)

\begin{table}
\begin{center}
\begin{tabular}{|c||c|c|}
\hline
Geometry	& Dispersion relation & $\tR$ \\
\hline
&& \\
~SAdS$_4$  (M2 \& D1)~
     & $~\displaystyle{ \nw = - \frac{3i \nq^2}{2} 
     - i\, \frac{3(9\ln3+\sqrt{3}\pi)}{16} \nq^4 + O(\nq^6) }~$
     & $~\displaystyle{\frac{9\ln3+\sqrt{3}\pi}{24\pi T} }~$ \\
&& \\
~SAdS$_5$ (D3)~
     & $~\displaystyle{ \nw = - i \nq^2 - i\, (\ln 2) \nq^4 + O(\nq^6) }~$
     & $~\displaystyle{\frac{\ln2}{2\pi T} }~$ \\
&& \\
~SAdS$_7$ (M5 \& D4)~
     & $~\displaystyle{ \nw = - \frac{3i \nq^2}{4} 
     - i\, \frac{3(9\ln3-\sqrt{3}\pi)}{64} \nq^4 + O(\nq^6) }~$
     & $~\displaystyle{\frac{9\ln3-\sqrt{3}\pi}{24\pi T} }~$ \\
&& \\
\hline
\end{tabular}
\caption{The relaxation time $\tR$ computed from the diffusive mode. 
}
\label{table:resultsI}
\vspace{5mm}
%
\begin{tabular}{|c||c|c|}
\hline
Geometry	& Dispersion relation & $\tau_\pi$ from shear mode \\
\hline
&& \\
~SAdS$_4$  (M2)~ 
     & $~\displaystyle{ \nw = - \frac{i \nq^2}{2} 
     - i\, \frac{9-9\ln3+\sqrt{3}\pi}{48} \nq^4 + O(\nq^6) }~$
     & $~\displaystyle{\frac{9-(9\ln3-\sqrt{3}\pi)}{24\pi T} }~$ \\
&& \\
~SAdS$_5$ (D3)~ 
     & $~\displaystyle{ \nw = - \frac{i \nq^2}{2} 
     - i\, \frac{1-\ln 2}{4} \nq^4 + O(\nq^6) }~$
     & $~\displaystyle{\frac{1-\ln 2}{2\pi T} }~$ \\
&& \\
~SAdS$_7$ (M5 \& D4)~ 
     & $~\displaystyle{ \nw = - \frac{i \nq^2}{2} 
     - i\, \frac{18-9\ln3-\sqrt{3}\pi}{48} \nq^4 + O(\nq^6) }~$
     & $~\displaystyle{\frac{18-(9\ln3+\sqrt{3}\pi)}{24\pi T} }~$ \\
&& \\
\hline
\end{tabular}
\caption{The relaxation time $\tau_\pi$ computed from the shear mode. 
}
\label{table:resultsII}
\vspace{5mm}
%
\begin{tabular}{|c||c|c|}
\hline
Geometry	&Dispersion relation &$\tpi$ from sound mode  \\
\hline
&& \\
~SAdS$_4$  (M2) ~
     & $~\displaystyle{ \nw = \frac{\nq}{\sqrt{2}} - \frac{i \nq^2}{4} 
     +\frac{15-9 \ln 3+\sqrt{3} \pi }{48 \sqrt{2}} \nq^3 + O(\nq^4) }~$
     & $~\displaystyle{\frac{18-(9\ln3-\sqrt{3}\pi)}{24\pi T} \sim 0.18\,\fm}~$ \\
     && \\
~SAdS$_5$ (D3)~
     & $~\displaystyle{ \nw = \frac{\nq}{\sqrt{3}} - \frac{i \nq^2}{3} 
     + \frac{3-2\ln 2}{6 \sqrt{3}} \nq^3 + O(\nq^4) }~$
     & $~\displaystyle{\frac{2-\ln 2}{2\pi T}  \sim 0.21\,\fm}~$ \\
     && \\
~SAdS$_7$ (M5)~
     & $~\displaystyle{ \nw = \frac{\nq}{\sqrt{5}} - \frac{2 i \nq^2}{5} 
     + \frac{24-9 \ln 3-\sqrt{3} \pi}{30 \sqrt{5}}\nq^3 + O(\nq^4) }~$
     & $~\displaystyle{\frac{36-(9\ln3+\sqrt{3}\pi)}{24\pi T}  \sim 0.27\,\fm}~$ \\
     && \\
\hline
\end{tabular}
\caption{The relaxation time $\tpi$ computed from the sound mode. Numerical values shown correspond to $T^{-1}=1$~fm.}
\label{table:resultsIII}
\end{center}
\end{table}

\subsection{The sound mode}

The sound mode is interesting in the sense that the relaxation time $\tpi$ appears in this mode as well. [See \eq{hydro-decouple-dipersion_rel-sound-conf}.] Since $\tau_\Pi=0$ for conformal theories, one can deduce $\tpi$ from this mode as well, but here one encounters a puzzle. From the sound mode of the ${\cal N}=4$ SYM, we get
\be
\nw= \frac{\nq}{\sqrt{3}} - \frac{i \nq^2}{3} + \frac{3-2\ln 2}{6\sqrt{3}} \nq^3 +O(\nq^4)~.
\ee
Comparing this with the dispersion relation (\ref{eq:hydro-decouple-dipersion_rel-sound-conf}), one obtains
\be
\tpi = \frac{2-\ln 2}{2\pi T}~,
\label{eq:tau_pi_sound}
\ee
which does not agree with the answer obtained from the shear mode (\ref{eq:tau_pi_vector}). The other results are summarized in \tabl{resultsIII}. One can see similar discrepancies for the other SAdS$_{p+2}$ backgrounds as well.

We have not located the origin of the problem. But, first of all, the Israel-Stewart theory is not the unique formalism for causal hydrodynamics. One well-known alternative is the ``divergence type theories" \cite{LMR,Geroch:1990bw}. The discrepancies we found may imply that the gauge theory plasmas do not really fit into the framework of the Israel-Stewart theory.%
\footnote{One interpretation is proposed in Ref.~\cite{Baier:2007ix}, which appeared just before we submitted the first version of this paper. According to the paper, the shear mode is unreliable to obtain $\tpi$, and one should use the sound mode. Thus, we will use \tabl{resultsII} for physical interpretations hereafter.}

For the D$p$-brane, the dispersion relation is identical to the one of the corresponding SAdS solution. However, the hydrodynamic interpretation is different partly due to the different spacetime dimensionality. They are nonconformal, so $\tPi \neq 0$ and one cannot determine $\tpi$ and $\tPi$ separately from the sound mode alone. For the D1-brane, 
\be
\tau_\Pi=\frac{18-(9\ln3-\sqrt{3}\pi)}{24\pi T}~
\ee
(No $\tpi$ for the D1-brane). For the D4-brane, using \eq{hydro-decouple-dipersion_rel-sound}, we get
\be
\tau_\pi+\frac{\tau_\Pi}{15} = 2\,\frac{36-(9\ln3+\sqrt{3}\pi)}{45\pi T} \sim 0.29\,\fm.
\ee


\subsection{Discussion}\label{sec:discussion}


The theories we consider here are not QCD. Thus, it is important to ask if there is any universality or any generic behaviors just like $\eta/s$. This is the reason why we consider various theories. From \tabl{resultsII}, there seems no obvious universality, but the numerical values of $\tpi$ are similar among the theories we consider. 

To be more specific, get some numbers. 
First, recall that $\hbar c \sim 197~{\rm MeV fm}$ and 197 MeV is not far from the QCD transition temperature $T_c$. This means that the characteristic length scale at $T_c$ is $T^{-1} \sim O(\fm)$, and this is the typical value one would expect for relaxation times. In fact, the kinetic theory predicts that \cite{Israel:1979wp}
\be
\tpi = \frac{3\eta}{2p} =\frac{6}{T}\frac{\eta}{s}
\label{eq:kinetic}
\ee
for a 4-dimensional Boltzmann gas, where we used $\epsilon+p=T s$ and used the fact that the ${\cal N}=4$ theory is conformal so that $\epsilon=3p$. If one uses the AdS/CFT value of $\eta/s=1/(4\pi)$, $ \tpi = 3/(2\pi T) \sim 0.5\, \fm $ for $T^{-1} = 1$ fm. 
This value is not far from the our results in \tabl{resultsIII}. 


This seems consistent with what Israel and Stewart found. They obtained \eq{kinetic} by analyzing the Boltzmann equation. More precisely, they estimated $\beta_2=\tpi/(2\eta)$.%
\footnote{$\beta_2$ is a parameter they use to parametrize causal hydrodynamics [\eq{def-tilQ}]. However, $\beta_2$ appears in the dispersion relation only in the combination $2\eta\beta_2$, which is $\tpi$. [\eq{def-tau_pi}]}
They found that $\beta_2$ is not sensitive to the value of the cross section. This implies that $\beta_2$ is more or less constant as we vary the coupling constant. Namely, $\eta$ strongly depends on the coupling, and so does $\tpi$, but $\tpi/\eta$ does not strongly depend on the coupling. (The entropy density $s$ does not strongly depend on the coupling \cite{Gubser:1996de}, so it is irrelevant here.) And in fact, we found that the ratio $\tpi/\eta$ from the AdS/CFT duality is not far from the kinetic theory estimate.

\begin{acknowledgments}
We would like to thank Kengo Maeda for discussions. We would especially like to thank Tetsufumi Hirano for various discussions and suggestions throughout this project.
\end{acknowledgments}

\vspace*{0.3cm}
{\bf Note added}:
While this paper is in preparation, a number of interesting papers appeared \cite{Benincasa:2007tp,Baier:2007ix,Bhattacharyya:2007jc}, which study the similar problem as ours (See also Ref.~\cite{Natsuume:2008iy}). In particular, Ref.~\cite{Baier:2007ix} partly uses the same technique as ours. For SAdS$_5$, our results of $\tpi$  coincide with the results of Ref.~\cite{Baier:2007ix} both in the shear mode and in the sound mode. Also, Ref.~\cite{Baier:2007ix} argues that the shear mode result is unreliable and one should use the sound mode to extract $\tpi$. Based on their observation, we have changed an early interpretation based only on the shear mode. 
See Ref.~\cite{causal_review} for a review. 

\appendix

\section{Israel-Stewart theory}\label{sec:IS}

Sections~\ref{sec:preliminaries} and \ref{sec:2nd_law} review the Israel-Stewart theory; Readers who are familiar with the formalism may go to \sect{dispersion} directly.

\subsection{Preliminaries}\label{sec:preliminaries}

Denote the number of {\it spatial} dimensions by $d_s$. We consider the case of only one conserved charge $\rho$ for simplicity, but the case of several charges is straightforward.
The fundamental variables in hydrodynamics are the conserved current $j^\mu$, the energy-momentum tensor $T^{\mu\nu}$, and the entropy current $s^\mu$ (which gives the direction of time).

\subsubsection{Equilibrium}

In equilibrium, there is a special ``fluid rest frame" defined by $u_{\rm eq}^\mu$ $(u_{\rm eq}^2=-1)$, in which there is no (spatial) flow. Thus,
\begin{align}
  & j^\mu = \rho\,u_{\rm eq}^\mu~,
& & T^{\mu\nu} = \epsilon\,u_{\rm eq}^\mu\, u_{\rm eq}^\nu
  + p_{\rm eq}\,( g^{\mu\nu} + u_{\rm eq}^\mu\, u_{\rm eq}^\nu )~,
& & s^\mu = s_{\rm eq}\,u_{\rm eq}^\mu~.
\label{eq:EQ_current}
\end{align}
The first law $ { T_{\rm eq} } ds_{\rm eq} = d\epsilon - \mu_{\rm eq} d\rho$
tells that $s_{\rm eq}$ is not an independent variable, but it is a function of $\epsilon$ and $\rho$.
Also,  the temperature $T_{\rm eq}$ and the chemical potential $\mu_{\rm eq}$ are defined by the first law as
\begin{align}
  & \frac{1}{ T_{\rm eq} }
  := \frac{\partial s_{\rm eq}(\epsilon, \rho)}{\partial \epsilon}~,
& & \frac{ \mu_{\rm eq} }{ T_{\rm eq} }
  := \frac{\partial s_{\rm eq}(\epsilon, \rho)}{\partial \rho}~.
\label{eq:def-T_eq-mu_eq}
\end{align}
The pressure $p_{\rm eq}$ is not independent either due to the Euler identity $p_{\rm eq} = - \epsilon + T_{\rm eq}\,s_{\rm eq} + \mu_{\rm eq}\,\rho$. 
It is convenient to rewrite the Euler identity in a covariant manner:
\be
   s^\mu
   = \frac{1}{ T_{\rm eq} }( p_{\rm eq}\, u_{\rm eq}^\mu
  - u_{{\rm eq},\nu}\, T^{\mu\nu}
  - \mu_{\rm eq}\, j^\mu)~.
\label{eq:EQ-Gibbs_rel-arrange}
\ee

\subsubsection{Near-equilibrium}

We closely follow Ref.~\cite{Israel:1976tn} but use slightly different conventions and notations. 
Consider a state of near-equilibrium whose deviation $\delta$ from the equilibrium is small. In equilibrium, the entropy density $s_{\rm eq}$ is a function of the charge $\rho$ and the energy density $\epsilon$. We assume that the entropy current $s^\mu$ is a function of the currents $j^\mu$ and $T^{\mu\nu}$ even in the case of a nonequilibrium state:
\begin{align*}
  & s^\mu = s^\mu( j^\mu, T^{\mu\nu} )~.
\end{align*}

For a nonequilibrium state, various currents have spatial flows and they do not match in general. Thus, the notion of the ``fluid rest frame" is ambiguous: a different current defines a different ``fluid rest frame." 
There are two common choices for the ``fluid rest frame" in the literature (The notations are defined below):
\begin{enumerate}
\item The Eckart frame or Particle frame (N-frame): $j_{\perp}^\mu=0$ in this frame.
\item The Landau-Lifshitz frame or Energy frame (E-frame): $k_\perp^\mu=0$ in this frame.
\end{enumerate}
Instead of choosing a particular frame, we consider a general reference frame $u^\mu$ $(u^2=-1)$ which is close to a fictitious rest frame of equilibrium thermodynamics, where
\begin{align}
  & \big\vert\, u^\mu - u_{\rm eq}^\mu\, \big\vert = O(\delta)~.
\end{align}
Then, we derive the results so that they do not depend on a choice of $u^\mu$ (``frame-invariance"). Namely, we use neither the Eckart frame nor the Landau-Lifshitz frame. (However, we frequently comment the case of the Landau-Lifshitz frame since it is frequently used.)

%
%

Given $u^\mu$, one naturally defines a $(d_s+1)$-decomposition of the spacetime tensor $g^{\mu\nu}$ by the projection operator $h^{\mu\nu}(u)$:
%
\begin{align}
  & h^{\mu\nu} := g^{\mu\nu} + u^\mu\, u^\nu~.
\label{eq:def-h_metric}
\end{align}
Then, $j^\mu$ and $T^{\mu\nu}$ are decomposed as
\begin{align}
  & j^{\mu} = \rho\,u^\mu + j_{\perp}^\mu~,
\label{eq:n_A-decompose} \\
  & T^{\mu\nu}
  = \epsilon\,u^\mu\, u^\nu + 2\, k_\perp^{(\mu}\, u^{\nu)}
  +(\hp + \Pi)\, h^{\mu\nu} + \pi^{\mu\nu}~.
\label{eq:T-decompose}
\end{align}
Here, the variables with ``$\perp$" represent the components which are orthogonal to $u^\mu$: {\it e.g.}, $u_\mu j_{\perp}^\mu = 0$.  
($\pi^{\mu\nu}$ also satisfies $u_\mu \pi^{\mu\nu} = 0$.)
The quantities with ``$\Hat{~}$" are the quantities defined by the functional form of the entropy density $s_{\rm eq}(\epsilon, \rho)$ {\it in the equilibrium}. For example, use \eq{def-T_eq-mu_eq} for $\hT(u)$ and $\hmu(u)$.
By construction, 
the ``net flow of charge" $j_{\perp}^\mu(u)$, 
the ``energy flow" $k_\perp^\mu(u)$,
the trace and traceless part of the viscous stress $\Pi(u)$ and $\pi^{\mu\nu}(u)$
should satisfy
\begin{align*}
  & \big\vert\, j_{\perp}^\mu\, \big\vert,~
  \big\vert\, k_\perp^\mu\, \big\vert,~
  \big\vert\, \Pi\, \big\vert,~
  \big\vert\, \pi^{\mu\nu}\, \big\vert = O(\delta)~.
\end{align*}

We will make use of the ``covariant time derivative" $\nabla_u := u^\mu\nabla_\mu$ and the ``covariant spatial derivative" $D^\mu$: {\it e.g.}, 
$D_\mu v_\perp^\nu = h_\mu{}^\rho h^\nu{}_\sigma
\nabla_\rho v_\perp^\sigma$
for a spatial vector $v_\perp^\mu$.

Also, one can write
\begin{align}
  & \nabla_\mu u_\nu
  = - u_\mu\, a_\nu + \frac{\theta}{d_s}\, h_{\mu\nu}
  + \sigma_{\mu\nu} + \omega_{\mu\nu}~,
\label{eq:nabla_u-decompose}
\end{align}
using
\begin{align}
  & \theta := \nabla_\mu u^\mu
  = h^\nu{}_\mu \nabla_\nu u^\mu~,
\label{eq:def-theta} \\
  & a^\mu := \nabla_u u^\mu~,
\label{eq:def-accelerate} \\
  & \sigma_{\mu\nu}
  := \left( h_{(\mu}{}^\rho\, h_{\nu)}{}^\sigma
  - \frac{ h_{\mu\nu}\, h^{\rho\sigma} }{d_s} \right)~
  \nabla_\rho u_\sigma~,
\label{eq:def-sigma} \\
  & \omega_{\mu\nu}
  := h_{[\mu}{}^\rho\, h_{\nu]}{}^\sigma~\nabla_\rho u_\sigma~.
\label{eq:def-omega}
\end{align}
These quantities represent the expansion, acceleration, shear, and rotation of the reference frame, respectively. 

%
%

Now, the fundamental variables $j^\mu$, $T^{\mu\nu}$, and $s^\mu$ do not depend on a particular frame, but their $(d_s+1)$-decompositions depend on a frame. Thus, let us check which variables are frame-invariant. Let us consider the following transformation of the reference frame:
\begin{align}
  & u^\mu \hspace{0.3cm} \longrightarrow \hspace{0.3cm}
  \bar{u}^\mu
  := ( 1 + \zeta_\perp^2 )^{1/2}\, u^\mu + \zeta_\perp^\mu~,
\label{eq:def-u_bar} \\
  & \text{where}\quad
  \bar{u}^2 = -1~, \quad
  u_\nu \zeta_\perp^\nu = 0~, \quad
  \big\vert\, \zeta_\perp^\nu\, \big\vert = O(\delta)~.\end{align}
One can check (See Appendix of Ref.~\cite{Israel:1976tn})
\begin{itemize}
\item The variations are $O(\delta^2)$:  $\rho$, $\epsilon$, $\Pi$, and $\pi^{\mu\nu}$.%
\footnote{Obviously, $\hp$, $\hmu$, and $\hs$ are also frame-invariant as well since they are defined through $\rho$ and $\epsilon$.}
\item The variations are $O(\delta)$: $j_{\perp}^\mu$ and $k_\perp^\mu$.
\end{itemize}
Thus, the former are the frame-invariant quantities up to $O(\delta)$. The variables $j_{\perp}^\mu$ and $k_\perp^\mu$ are not, but they combine to give  frame-invariant quantities up to $O(\delta)$: 
\begin{align}
  & \invj^\mu
  := j_{\perp}^\mu - \frac{\rho}{\epsilon + \hp}\, k_\perp^\mu~,
\label{eq:def-nu_A} \\
  & \invu^\mu
  := u^\mu + \frac{k_\perp^\mu}{\epsilon + \hp}~.
\label{eq:def-u_E}
\end{align}
Note that $\invj^\mu$ consists partly of $j_\perp^\mu$ and partly of $k_\perp^\mu$. 
In the Landau-Lifshitz frame where $k_\perp^\mu=0$, $\invj^\mu = j^\mu$ and $\invu^\mu = u^\mu$. Using these variables, one can rewrite the currents as
\begin{align}
  & j^{\mu} = \rho\,\invu^\mu + \invj^\mu~,
\label{eq:n_A-decompose-II} \\
  & T^{\mu\nu}
  = \big( \epsilon + \hp \big)\,\invu^\mu\, \invu^\nu
  - \frac{ k_\perp^\mu\, k_\perp^\nu}{\epsilon + \hp} 
  + \hp\, g^{\mu\nu} + \Pi\, h^{\mu\nu}
  + \pi^{\mu\nu}~.
\label{eq:T-decompose-II}
\end{align}
%

\subsection{Entropy and the second law of thermodynamics}\label{sec:2nd_law}

The constitutive equations are constructed so that the second law of thermodynamics $\nabla_\mu s^\mu \ge 0$ is guaranteed.
Therefore, the form of the entropy current $s^\mu$ (as a function of $j^\mu$ and $T^{\mu\nu}$) becomes important. 

Assume that the Euler identity (\ref{eq:EQ-Gibbs_rel-arrange}) remains a good approximation even for nonequilibrium states. Thus, define $Q^\mu(u)$ by
\begin{align}
   s^\mu
  &=: \frac{1}{\hT}\, \left( \hp\, u^\mu - u_\nu\, T^{\mu\nu}
  - \hmu\, j^\mu \right) - Q^\mu~,
\label{eq:entropy_current-decompose}
\end{align}
where $\big\vert Q^\mu \big\vert = O(\delta)$.

Fro small deviations, it suffices to retain only the $O(\delta^2)$ terms for $Q^\mu$. We take the form%
\footnote{Since we used frame-invariant quantities, it is clear that $\tilQ$ is frame-invariant up to $O(\delta^2)$. It is convenient to construct $\tilQ$ in this way since $\nabla_\mu s^\mu = O(\delta^2)$. On the other hand, $R^\mu$ is not frame-invariant. The term $R^\mu$ is chosen such that it cancels unphysical $O(\delta)$ terms in $\nabla_\mu s^\mu$, which appear in the ``first order formalism" below.}
\begin{align}
  & Q^\mu
  =: \tilQ^\mu + R^\mu~,
\label{eq:def-Q_MIS}
\end{align}
where
\begin{align}
   \tilQ^\mu
  &= \frac{u^\mu}{2\, \hT} \left( \beta_0\,\Pi^2
  + \beta_1\, \invj \cdot \invj
  + \beta_2\, \pi^{\rho\sigma} \pi_{\rho\sigma} \right)
  + \frac{\alpha_0}{\hT}\,\Pi\,\invj^\mu
  + \frac{\alpha_1}{\hT}\,
    \pi^{\mu\lambda}\, \invj_{\lambda}~,
\label{eq:def-tilQ} \\
   R^\mu
  &:= \frac{1}{\hT\, ( \epsilon + \hp )} \left(
    \pi^{\mu\lambda}\, k_{\perp,\lambda} + \Pi\, k_\perp^\mu
  + \frac{u^\mu}{2}\,k_\perp \cdot k_\perp \right)~.
\label{eq:def-R}
\end{align}
Note that $R^\mu$ vanishes in the Landau-Lifshitz frame. Five constants $\alpha_A$ and $\beta_A$ appeared in $Q^\mu$: $\beta_A$ represent the relaxation times and $\alpha_A$ represent the couplings among various modes as we will see below.

\subsubsection{``First order formalism"}\label{sec:1st}

The divergence of Eq.~(\ref{eq:entropy_current-decompose}) is given by
\be
   \nabla_\mu s^\mu
  = - \frac{ \pi^{\mu\nu} + \Pi\, h^{\mu\nu} }{\hT}\, \nabla_\mu\, \invu_\nu
  - \invj_{\mu}\, D^\mu \left( \frac{\hmu}{\hT} \right)
  - \nabla_\mu \tilQ^\mu
  + O(\delta^3)~.
\label{eq:entropy_current-divergence-IV}
\ee
Here, we have taken $R^\mu$ into account (See \sect{2nd}).
In order to ensure the second law of thermodynamics, the right-hand side of Eq.~(\ref{eq:entropy_current-divergence-IV}) must be positive-definite: This is how the constitutive equations are derived.
When one ignores $\tilQ^\mu$ (``first order formalism"), $\nabla_\mu s^\mu \ge 0$ is guaranteed if the right-hand side is a sum of complete squares. Thus, introducing the transport coefficients,
$D$, $\zeta$, and $\eta$, we require
\begin{align}
  & \invj^\lambda
  = - D\, \hT\,
  \left( \frac{\partial \hmu}{\partial \rho} \right)_{\hT}^{-1}\, 
  D^\lambda \left( \frac{\hmu}{\hT} \right)~,
\label{eq:number_current-first-I} \\
  & \Pi = - \zeta\, \Theta~,
\label{eq:bulk_viscosity-first-I} \\
  & \pi^{\mu\nu} = - 2\, \eta\, \Sigma^{\mu\nu}~,
\label{eq:shear_viscosity-first-I}
\end{align}
where
%
\begin{align}
  & \Theta
  := h^{\mu\nu}\, \nabla_\mu \invu_\nu~,
  \qquad
  \Sigma_{\mu\nu}
  := \left( h_{(\mu}{}^\rho\, h_{\nu)}{}^\sigma
  - \frac{h_{\mu\nu}\, h^{\rho\sigma}}{d_s} \right) \, \nabla_\rho \invu_\sigma~.
\end{align}
The variables $\Theta$ and $ \Sigma_{\mu\nu}$ are the frame-invariant expansion and shear, respectively.
Then,
\begin{align}
   \nabla_\mu s^\mu
  &= \frac{\invj^2}{D\, \hT}\,
    \left( \frac{\partial \hmu}{\partial \rho} \right)_{\hT}
  + \frac{\Pi^2}{\zeta\, \hT}
  + \frac{\pi^{\mu\nu} \pi_{\mu\nu}}{2\, \eta\, \hT}
  - \nabla_\mu \Tilde{Q}^\mu
  + O(\delta^3)~,
\label{eq:entropy_current-divergence-first}
\end{align}
so $\nabla_\mu s^\mu \ge 0$ is ensured if $D > 0, \zeta > 0$, and $\eta > 0$, provided that $( \partial \hmu/\partial \rho )_{\hT}>0$.

Note that the stress tensor is rewritten as
\begin{align}
  & T^{ij}
  = \hp\, h^{ij} - \zeta\, \Theta\, h^{ij}
  - 2\, \eta\, \Sigma^{ij}~,
\label{eq:T_ij-decompose}
\end{align}
using Eqs.~(\ref{eq:bulk_viscosity-first-I}) and (\ref{eq:shear_viscosity-first-I}). This is just the familiar form for the stress tensor.

\subsubsection{Second order formalism}\label{sec:2nd}

``The first order formalism" is not a closed form since one has to take into account the $O(\delta^2)$ terms $R^\mu$.%
\footnote{The Landau-Lifshitz frame is free of this problem since $R^\mu=0$.}
If one does not include this term, $\nabla_\mu s^\mu$ becomes frame-dependent; Moreover, hydrodynamic equations become an overdetermined system. On the other hand, if one includes this term as was done in \sect{1st}, one had better include all $O(\delta^2)$ terms in $Q^\mu$. The second order formalism takes $\tilQ^\mu$ into account. In this case, the divergence of the entropy current gives
\begin{align}
   \nabla_\mu s^\mu
  &= - \frac{ \pi_{\mu\lambda} }{\hT}\, \left( \Sigma^{\mu\lambda}
  + \beta_2~\nabla_\invu \pi^{\mu\lambda}
  + \alpha_1~\nabla^\mu \invj^\lambda \right)
   - \frac{\Pi}{\hT}\, \left( \Theta
  + \beta_0\, \nabla_\invu \Pi
  + \alpha_0\, \nabla_\mu \invj^\mu \right)
\nonumber \\
  &- \frac{ \invj_{\mu} }{\hT}\,
            \left( \hT\, D^\mu \left(\frac{\hmu}{\hT}\right)
  + \beta_1\, 
      \nabla_\invu \invj^\mu
  + \alpha_0\, \nabla^\mu \Pi
  + \alpha_1\,  \nabla_\nu \pi^{\mu\nu} \right)
  + O(\delta^3)~,
\label{eq:entropy_current-divergence-V}
\end{align}
where $\nabla_\invu := \invu^\mu\nabla_\mu$. 
Then, the constitutive equations for the second order formalism are given by
\begin{align}
  & \invj^\lambda
  = - D\,
  \left( \frac{\partial \hmu}{\partial \rho} \right)_{\hT}^{-1}\, 
  \Bigg[~\hT\, D^\lambda \left( \frac{\hmu}{\hT} \right)
  + \beta_1\, h^\lambda{}_\rho\,
     \nabla_\invu \invj^\rho
  + \alpha_0\, D^\lambda \Pi
  + \alpha_1\, \big( D_\nu \pi^{\nu\lambda}
  + \pi^{\lambda\nu}\, a_\nu \big)~\Bigg]~,
\label{eq:number_current-second-I} \\
  & \Pi
  = - \zeta\, \left( \Theta
  + \beta_0\, \nabla_\invu \Pi
  + \alpha_0\, \nabla_\mu \invj^\mu \right)~,
\label{eq:bulk_viscosity-second-I} \\
  & \pi^{\mu\lambda}
  = - 2\, \eta\, \left[~\Sigma^{\mu\lambda}
  + \left( h^\mu{}_\rho\, h^\lambda{}_\sigma
  - \frac{h^{\mu\lambda}\, h_{\rho\sigma}}{d_s} \right)
    \left( \beta_2\, \nabla_\invu \pi^{\rho\sigma}
  + \alpha_1\, D^{(\rho} \invj^{\sigma)} \right)~
  \right]~.
\label{eq:shear_viscosity-second-I}
\end{align}
%

\subsection{Dispersion relations}\label{sec:dispersion}

\subsubsection{Assumptions and tensor decomposition}

In order to obtain the dispersion relations for causal hydrodynamics, we make a number of simplifying assumptions:
\begin{itemize}
\item {\it Linear perturbations}: We consider linear perturbations from the thermal equilibrium.
\item {\it Rest frame}: We choose the fluid rest frame in equilibrium as the reference frame $u^\mu$ so that 
\be
\theta = 0~, \qquad \sigma_{\mu\nu} = \omega_{\mu\nu} = 0~.
\ee
Moreover, all thermodynamic quantities have no time-dependence and there are no flows, so
\begin{align*}
  & \bm{\invj}^\lambda = \bmP_\perp^\mu = 0~,
& & \bm{\invu}^\mu = u^\mu~,
& & \delta \invu^\mu\, u_\mu = 0~.
\end{align*}
(The boldface letters represent background values 
and $\delta U^\mu := U^\mu - \bmU^\mu$.)
\item {\it Flat (boundary) spacetime}: We consider the flat $(d_s+1)$-dimensional spacetime. Then, $a^\mu=0$.
\item {\it ``Decoupled ansatz"}: Let us take into account
  the {\it bulk results in advance}.
  In the text,
  we consider the backgrounds with no R-charge, so
  one can set $\bmn=\Hat{\bmmu}=0$.
  Also, it holds
  $(\partial \Hat{\bmp}/\partial \bmn)_{\bmeps}
  = (\partial (\Hat{\bmmu}\, \Hat{\bmT}^{-1})/\partial \bmeps)_{\bmn}
  = 0$, and
  the gauge field and the metric perturbations
  decouple. This implies $\alpha_0=\alpha_1=0$.

\end{itemize}

Introduce the time coordinate by $u^\mu = \big( \partial_t \big)^\mu$, and denote spatial indices as $i, j, k, \cdots$. The linear perturbations are defined, {\it e.g.}, by $\rho =: \bmn+\delta\rho$.
Using Eqs.~(\ref{eq:n_A-decompose-II}) and (\ref{eq:T-decompose-II}), one gets the conservation equations 
\begin{align}
  & 0
  = \partial_t \delta \rho
  + \bmn\, \delta \Theta
  + D_i \delta\invj^i~,
\label{eq:pert-EOM-n_A-flat-II} \\
  & 0
  = \partial_t \delta \epsilon
  + \big( \bmeps + \Hat{\bmp} \big)\, \delta \Theta~,
\label{eq:pert-EOM-EM_tensor-flat_time-II} \\
  & 0
  = \partial_t \delta \invu^i
  + \frac{ D^i \big( \delta \hp + \delta \Pi \big)
         + D_j \delta \pi^{ij} }{ \bmeps + \Hat{\bmp} }~.
\label{eq:pert-EOM-EM_tensor-flat_space-II}
\end{align}
Using Eqs.~(\ref{eq:number_current-second-I})-(\ref{eq:shear_viscosity-second-I}), one gets the constitutive equations 
\begin{align}
  & \delta \invj^i
  = - D\,
  \left( \frac{\partial \Hat{\bmmu}}{\partial \bmn} \right)_{\Hat{\bmT}}^{-1}\, 
  \Bigg[~\Hat{\bmT}\,
    D^i \delta \left( \frac{\hmu}{\hT} \right)
  + \beta_1\, \partial_t \delta \invj^i
  + \alpha_0\, D^i \delta \Pi
  + \alpha_1 D_j \delta \pi^{ij}~\Bigg]~,
\label{eq:pert-number_current-second-II} \\
  & \delta \Pi
  = - \zeta\, \left[~\delta\Theta
  + \beta_0\, \partial_t \delta\Pi
  + \alpha_0\, D_j \delta\invj^j~\right]~,
\label{eq:pert-bulk_viscosity-second-II} \\
  & \delta \pi^{ij}
  = - 2\, \eta\, \left[~\delta\Sigma^{ij}
  + \left( h^i{}_k\, h^j{}_l - \frac{h^{ij}\, h_{kl}}{d_s} \right)
    \left( \beta_2\, \partial_t \delta\pi^{kl}
  + \alpha_1\, D^{(k} \delta\invj^{l)} \right)~
  \right]~.
\label{eq:pert-shear_viscosity-second-II}
\end{align}
We henceforth use the ``decoupled ansatz."

In order to solve these equations (\ref{eq:pert-EOM-n_A-flat-II})-(\ref{eq:pert-shear_viscosity-second-II}), decompose the spatial tensors as follows ($\Delta :=D_j D^j$):
\begin{align}
  & \delta \invj^i
  = D^i \invj_{L} + \invj_{T}^i~,
\label{eq:tensor_decomp-nu} \\
  & \delta \invu^i
  = D^i \invu_{L} + \invu_{T}^i~,
\label{eq:tensor_decomp-U} \\
  & \delta \pi^{ij}
  = \pi_L\, h^{ij}
  + \left( D^i D^j - \frac{h^{ij}}{d_s}\, \Delta \right)\,
    \pi_T^{} + 2\, D^{(i} \pi_T^{j)} + \pi_T^{ij}~.
\label{eq:tensor_decomp-pi}
\end{align}
Here,  $\invj_{T}^{i}$, $\invu_{T}^{i}$, $\pi_T^{i}$, and $\pi_T^{ij}$ represent the transverse components which satisfy
\begin{align}
  & D_i \invj_{T}^{i} = 0~,
& & D_i \invu_{T}^{i} = 0~,
& & D_i \pi_T^{i} = 0~,
\label{eq:transverse_cond-I} \\
  & D_j \pi_T^{ij} = 0~,
& & \pi_T^{~i}{}_i = 0~.
\label{eq:transverse_cond-II}
\end{align}
Note that $\pi_L = 0$ since $\delta \pi^{ij}$ is traceless.

\subsubsection{Vector modes}

The vector modes consist of  $\invu_{T}^{i}$, $\invj_{T}^{i}$,
and $\pi_{T}^{i}$. The equations for these fields are given by
%
\begin{align}
  & 0
  = \partial_t \invu_T^{i}
  + \frac{ \Delta \pi_T^{i} }{ \bmeps + \Hat{\bmp} }~,
\label{eq:vector_mode-EOM-I} \\
  & 0
  = \tR\, \partial_t \invj_{T}^{i} + \invj_{T}^{i}~,
\label{eq:vector_mode-EOM-II} \\
  & 0
  = \tpi\, \partial_t \pi_T^{i}
  + \pi_T^{i} + \eta\, \invu_T^{i}~,
\label{eq:vector_mode-EOM-III}
\end{align}
where the relaxation times are defined by 
\begin{align}
  & \tpi := 2\, \eta\, \beta_2~,
\label{eq:def-tau_pi} \\
  & \tR
  := D\,
  \left( \frac{\partial \Hat{\bmmu}}{\partial \bmn} \right)_{\Hat{\bmT}}^{-1}\, 
    \beta_1~.
\label{eq:def-tau_nu} 
\end{align}
%
%
%
Eqs.~(\ref{eq:vector_mode-EOM-I}) and (\ref{eq:vector_mode-EOM-III}) combine to give
\be
0  = \tpi\, \partial_t^2 \pi_T^{i}
  + \partial_t \pi_T^{i}
  - \frac{\eta}{ \bmeps + \Hat{\bmp} }\, \Delta \pi_T^{i}~.
\label{eq:decouple-vector_mode-EOM}
\ee
%
%
The Fourier-Laplace transform of the equation
\begin{align}
  & f(w, q)
  := \int^\infty_0 dt \int dz\, 
    e^{- i w t + iqz }\, f(t, z)~,
\label{eq:def-FL_transform}
\end{align}
gives 
\begin{align}
  & 0
  = - \tpi\,w^2
  -i\, w + \frac{\eta\, q^2}{\Hat{\bmT}\, \Hat{\bms}}~.
\label{eq:normalized-decouple-dipersion_rel-vector}
\end{align}
Here, we used the Euler identity
$\bmeps + \Hat{\bmp} = \Hat{\bmT}\, \Hat{\bms}$ for zero chemical potential.

Expanding \eq{normalized-decouple-dipersion_rel-vector} for small $w$ and $q^2$, one obtains
\be
w = - i\,D_\eta\, q^2 - i\,D_\eta^2\,\tpi\,q^4 + O(q^6)~,
\label{eq:hydro-decouple-dipersion_rel-vector}
\ee
where $D_\eta := \eta/(\Hat{\bmT}\, \Hat{\bms})$. 

\subsubsection{Scalar mode (diffusive mode)}

There are two scalar modes: the diffusive mode and the sound mode. We discuss them separately since they decouple due to the ``decoupled ansatz." 

%
%

The diffusive mode consists of $\delta \rho$ and $\invj_L$. 
One gets
\begin{align}
   0
  &= \partial_t \delta \rho + \Delta \invj_{L}~,
\label{eq:decouple-R_charge-EOM-I} \\
   0
  &= \tR\, \partial_t \invj_{L} + \invj_{L}
  + D\,
  \left( \frac{\partial \Hat{\bmmu}}{\partial \bmn} \right)_{\Hat{\bmT}}^{-1}\, 
  \Hat{\bmT}\, \left( \frac{\partial}{\partial \bmn}\,
        \frac{ \Hat{\bmmu} }{ \Hat{\bmT} } \right)_{\bmeps}\, \delta \rho
\nonumber \\
  &= \tR\, \partial_t \invj_{L} + \invj_{L} + D\, \delta \rho~,
\label{eq:decouple-R_charge-EOM-II} 
\end{align}
which combine to give
\be
   0 = \tR\, \partial_t^2 \delta \rho + \partial_t \delta \rho - D\, \Delta\, \delta \rho~.
\label{eq:decouple-R_charge-EOM-III}
\ee

This takes the same form as the shear mode equation (\ref{eq:decouple-vector_mode-EOM}), so one can immediately write the dispersion relation:
\be  
 0 = - \tR\, w^2 - i\, w + D\, q^2~,
\label{eq:decouple-dipersion_rel-charge}
\ee
or
\be
w = - i\, D\, q^2 - i\, D^2\,\tR\,q^4 + O(q^6)~.
\label{eq:hydro-decouple-dipersion_rel-charge}
\ee
%

\subsubsection{Scalar mode (sound mode)}

The sound mode consists of $\delta \epsilon$, $\invu_L$, $\delta \Pi$, and $\pi_T$.
One gets
\begin{align}
   0
  &= \partial_t \delta \epsilon
  + \big( \bmeps + \Hat{\bmp} \big)\, \Delta U_L~,
\label{eq:decouple-scalar_mode-EOM-I} \\
   0
  &= \partial_t U_L
  + \frac{ v_s^2\, \delta \epsilon + \delta \Pi
          + ( 1 - 1/d_s )\, \Delta \pi_T }{ \bmeps + \Hat{\bmp} }~,
\label{eq:decouple-scalar_mode-EOM-II} \\
   0
  &= \tau_\Pi~\partial_t \delta\Pi + \delta \Pi + \zeta~\Delta U_L~,
\label{eq:decouple-scalar_mode-EOM-III} \\
   0
  &= \tau_\pi~\partial_t \pi_T + \pi_T + 2\, \eta~U_L~,
\label{eq:decouple-scalar_mode-EOM-IV}
\end{align}
where
\bea
  v_s^2 &:=& \frac{\partial \Hat{\bmp}}{\partial \bmeps}~,\\
\label{eq:def-sound_velocity}
  \tau_\Pi &:=& \zeta\, \beta_0~.
\label{eq:def-tau_Pi} 
\eea

The Fourier-Laplace transformation of Eqs.~(\ref{eq:decouple-scalar_mode-EOM-I})-(\ref{eq:decouple-scalar_mode-EOM-IV}) gives
%
%
\begin{align}
  & 0
  = ( 1 - i\, \tau_\Pi\, w ) ( 1 - i\, \tau_\pi\, w )
    ( w^2 - v_s^2\, q^2 )
\nonumber \\
  &\hspace{1.0cm}
  + \frac{i\, w\, q^2}{ \bmeps + \Hat{\bmp} }~
    \left[~\zeta + 2\, \eta\, \left( 1 - \frac{1}{d_s} \right)
    - i\, w~\left\{ \zeta\, \tau_\pi
         + 2\, \left( 1 - \frac{1}{d_s} \right)\, \eta\, \tau_\Pi
     \right\}~\right]~.
\label{eq:decouple-dipersion_rel-sound} 
\end{align}
Thus, the dispersion relation for the hydrodynamic pole is
\bea
w &=&
\pm v_s\, q 
- \frac{i}{\bmeps + \Hat{\bmp}} \left(\frac{d_s -1}{d_s}\eta+\frac{\zeta}{2} \right) q^2 
\label{eq:hydro-decouple-dipersion_rel-sound}
\\
&& 
\pm \frac{1}{2 v_s (\bmeps + \Hat{\bmp})}
\left\{
   \frac{d_s -1}{d_s}\, \eta  
   \left( 2v_s^2 \tau_\pi - \frac{1-1/d_s}{\bmeps + \Hat{\bmp}}\, \eta
    \right)
   +\zeta
   \left( 
      v_s^2 \tau_\Pi - \frac{1-1/d_s}{\bmeps + \Hat{\bmp}}\, \eta 
      - \frac{\zeta}{4 (\bmeps + \Hat{\bmp})} 
    \right)
\right\} q^3 +\cdots~.
\nonumber 
\eea
In particular, for conformal theories where $\zeta = \tau_\Pi = 0$,
\begin{align}
   w
  &= \pm v_s\, q
  - i\, \frac{1-1/d_s}{ \bmeps + \Hat{\bmp} }~\eta\, q^2
  \pm \frac{1}{2\, v_s}\,
   \frac{1-1/d_s}{ \bmeps + \Hat{\bmp} }\, \eta
  \left( 2\, v_s^2\, \tau_\pi - \frac{1-1/d_s}{ \bmeps + \Hat{\bmp} }\, \eta
   \right) q^3 +O(q^4)~.
\label{eq:hydro-decouple-dipersion_rel-sound-conf}
\end{align}
%

\subsubsection{Tensor mode}

The tensor mode consists only of $\pi_T^{ij}$, and its equation is given by
\begin{align}
  & \tau_\pi~\partial_t \pi_T^{ij} + \pi_T^{ij} = 0~.
\label{eq:tensor_mode-EOM} \end{align}
Thus, the dispersion relation for the tensor mode is
\begin{align}
  & w + \frac{i}{\tau_\pi} = 0~,
\label{eq:dispersion_rel-tensor}
\end{align}
but this is unreliable since it is inconsistent with the hydrodynamic limit.

\section{Perturbative solutions for ${\cal N}=4$ SYM}\label{sec:solutions}

Here we give explicit expressions for the perturbative solutions for the ${\cal N}=4$ SYM. Integration constants are fixed by requiring the solutions to be regular at the horizon. For the shear mode,%
\footnote{
{\it Note added in v6:} 
In previous versions, we erroneously listed the solution for {\it the diffusive mode}.
However, $F_1$ as the diffusive mode solution is still incorrect and should be $F_1 = C \ln(1+u)/(2u)$.
}
\be
F_0 = C~, \qquad 
F_1 = i C \left(-\frac{1}{u}+\ln(1+u)\right)~, 
\qquad G_1 = \frac{C}{2u}~, 
\ee
\begin{align}
& F_2 =
C \left[~\frac{-1+\ln 2}{u}
+\frac{\ln 2}{2}\,  \ln (u-1)
-\frac{1}{4}\, \ln(u+1)\, \ln \Big\{ (1-u)^2 (u+1) \Big\}
-\frac{1}{2}\, \text{Li}_2 \left(\frac{u+1}{2}\right)\right]~,
\\
& H_{11} =
-i C \left[~\frac{1}{u}\, \left(1-\frac{1}{2} \ln (u+1)\right)
+\ln \left(\frac{u}{u+1}\right)\right]~,
\\
& G_2 =
\frac{1}{2}\, C\, \left[~\frac{1}{u}+\ln \left(\frac{u}{u+1}\right)\right]~,
\end{align}
where $\text{Li}_2(u)$ is a polylogarithm.

For the sound mode,
\be
F_0 = \frac{1}{4}C_1\left(1+\frac{1-3\,d_0^2}{u^2}\right)~, 
\ee
and the Dirichlet boundary condition at $u=0$ determines $d_0=\pm 1/\sqrt{3}$. Using these, we get
\be
F_1 = \mp \frac{C_1 (3\,d_1+i)}{2 \sqrt{3} u^2}~,
\ee
and the boundary condition gives $d_1=-i/3$. Finally, using these lower order results, we get
\bea
F_2 &=& 
    -\frac{1}{48 u^2} C \left[-\ln ^2(u+1) u^2+2\ln2 \ln (u-1) u^2
    -2 \ln (1-u) \ln (u+1) u^2+4\ln2 \ln (u+1) u^2 \right.
\nonumber \\
&& \left. 
    -2  \text{Li}_2\left(\frac{u+1}{2}\right) u^2+8 u \pm 24 \sqrt{3}\, d_2
    -8 \ln (u+1)+8\ln 2-12\right]~,
\eea
and the boundary condition gives $d_2=\pm(3-2\ln2)/6\sqrt{3}$.

\end{document}